\documentclass[]{aa}
\usepackage{natbib}
\usepackage{psfig}
\usepackage{wasysym}
\usepackage{txfonts}
\def\EBV{\mbox{E$_{\rm B-V}$}}

\def\AV{\mbox{A$_{\rm V}$}}

\def\sz{\mbox{${\sigma}_{\rm z}$}}
\def\nH2{\mbox{${\rm n}_\HH}$}

\def\pccc{~{\rm cm}^{-3}} 
 
\def\pcc {~{\rm cm}^{-2}}
\def\Tstar {\mbox{${\rm T}_{\rm r}^*$}}
\def\Tsub#1 {\mbox{${\rm T}_{\rm #1}$}}
\def\TK  {\Tsub K }

\def\arcsec{\mbox{$^{\prime\prime}$}} \def\arcmin{\mbox{$^{\prime}$}}

\def\degr{$^{\rm o}$}
\def\p{\mbox{$^+$}}

\def\h13cop{\mbox{{H$^{13}$CO\p}}}

\def\c3h2{\mbox{C$_3$H$_2$}}
 
 \def\R0{R$_0$}

\def\ddeg{{}^\circ\kern-.1em}

\def\kms{\mbox{km\,s$^{-1}$}}

\def\E#1 {$10^{#1}$}
\def\E#1 {E{#1}}
\def\P#1,{$\nH2\TK~=~#1\times~10^4\pccc$~K}
\def\ec#1,#2,#3,{#1\,(#2)\E{#3}}

\def\zoph{$\zeta$ Oph}

\def\H3{\mbox{H$_3$}}

\def\RH2{\mbox{R$_{\rm G}$}}
\def\g13{\mbox{g$_{13}$}} 

\newcommand{\emm}[1]{\ensuremath{#1}}   
\newcommand{\emr}[1]{\emm{\mathrm{#1}}} 
\newcommand{\paren}[1]{\emm{\left(  #1 \right) }} 

\newcommand{\mean}[1]{\emm{ \left<  #1 \right> }}
\newcommand{\abs}[1]{\emm{\left|  #1 \right| }} 

\newcommand{\hcop}{\emr{HCO^+}} 
\newcommand{\HI}{\emr{H\mathsc{i}}} 
\newcommand{\HH}{\emr{H_2}}
\newcommand{\cotw}{\emr{^{12}CO}}
\renewcommand{\coth}{\emr{^{13}CO}}
\newcommand{\coei}{\emr{C^{18}O}}

\newcommand{\N}[1]{\emr{N_{#1}}}
\newcommand{\NH}{\N{H}}
\newcommand{\NHI}{\N{\HI}}
\newcommand{\NHH}{\N{H_2}}
\newcommand{\NHCOp}{\N{\hcop}}
\newcommand{\NCO}{\N{CO}}
\newcommand{\NCH}{\N{CH}}

\newcommand{\X}[1]{\emm{X_\emr{#1}}}
\newcommand{\XCH}{\X{CH}}
\newcommand{\XCO}{\X{CO}}
\newcommand{\XOH}{\X{OH}}
\newcommand{\XHCOp}{\X{\hcop}}

\newcommand{\NhhWco}{\emm{\chi_\emr{CO}}}

\newcommand{\W}[1]{\emm{{\rm W}_\emr{#1}}}
\newcommand{\WCO}{\W{CO}}
\newcommand{\WCOperp}{\W{CO_\perp}}
\newcommand{\ZWCO}{\W{CO_0}}

\newcommand{\mytau}[1]{\emm{\int \tau(\emr{#1})\,dv}}
\newcommand{\tauhi}{\mytau{\HI}}
\newcommand{\tauhcop}{\mytau{\hcop}}

\newcommand{\f}[1]{\emm{f_\emr{#1}}}
\newcommand{\fHI}{\f{\HI}}
\newcommand{\fHH}{\f{\HH}}

\newcommand{\Kkms}{\emr{\,K\,km\,s^{-1}}}

\newcommand{\Rsun}{\emm{R_\odot}}

\title{The CO luminosity and CO-\HH\ conversion factor of diffuse ISM:
does CO emission trace dense molecular gas?}

\author{H. S. Liszt\inst{1}, J. Pety\inst{2,3} and R. Lucas \inst{4}}
\institute{National Radio Astronomy Observatory,
           520 Edgemont Road,
           Charlottesville, VA,
           USA 22903-2475
\and       Institut de Radioastronomie Millim\'etrique,
           300 Rue de la Piscine,
           F-38406 Saint Martin d'H\`eres,
           France
\and       Obs. de Paris, 
           61 av. de l'Observatoire, 75014, Paris, 
           France
\and       Al-MA, 
           Avda. Apoquindo 3846
           Piso 19, Edificio Alsacia, Las Condes,
           Santiago, Chile }

\begin{document}
\date{received \today}%
\offprints{H. S. Liszt}%
\mail{hliszt@nrao.edu}%
%
\abstract
{}
{We wish to separate and quantify the CO luminosity and CO-\HH\ conversion
  factor applicable to diffuse but partially-molecular ISM when \HH\ and CO
  are present but C\p\ is the dominant form of gas-phase carbon.}
{We discuss galactic lines of sight observed in \HI, \hcop\ and CO where CO
  emission is present but the intervening clouds are diffuse (locally \AV\ 
  $\la 1$ mag) with relatively small CO column densities $\NCO \la
  2\times10^{16}\pcc$.  We separate the atomic and molecular fractions
  statistically using \EBV\ as a gauge of the total gas column
  density and compare \NHH{} to the observed CO brightness.}
{Although there are \HH-bearing regions where CO emission is too faint to
  be detected, the mean ratio of integrated CO brightness to \NHH{} for
  diffuse ISM does not differ from the usual value of 1\Kkms\ of integrated
  CO brightness per $2\times10^{20}$ \HH $\pcc$ .  Moreover, the luminosity
  of diffuse CO viewed perpendicular to the galactic plane is 2/3 that seen
  at the Solar galactic radius in surveys of CO emission near the
  galactic plane.}
{Commonality of the CO-\HH\ conversion factors in diffuse and dark clouds
  can be understood from considerations of radiative transfer and CO
  chemistry.  There is unavoidable confusion between CO emission from
  diffuse and dark gas and misattribution of CO emission from diffuse to
  dark or giant molecular clouds.  The character of the ISM is different
  from what has been believed if CO and \HH\ that have been attributed to
  molecular clouds on the verge of star formation are actually in more
  tenuous, gravitationally-unbound diffuse gas.}

\authorrunning{Liszt, Pety \& Lucas} \titlerunning{CO luminosity of diffuse
  gas}

\maketitle{}

%

\section{Introduction}

It is a truism that sky maps of CO emission are understood as uniquely
tracing the Galaxy's molecular clouds, dense, cold strongly-shielded
regions where the hydrogen is predominantly molecular and the dominant form
of gas phase carbon is CO. Moreover, CO emission plays an
especially important role in ISM studies as the tracer of cold molecular
hydrogen through the use of the so-called CO-\HH\ conversion factor which
directly scales the integrated \cotw\ J=1-0 brightness \WCO{} to
\HH\ column density \NHH{}.  This deceptively simple conversion
is critically important to determining molecular and total gas column
densities and so to understanding the most basic properties of star
formation \citep{LerWal+08,BigLer+08,BotKen+09}, the origins of galactic dust
emission \citep{DraDal+07}, and other such fundamentals.

Yet, it is increasingly recognized that CO emission is present along lines
of sight lacking high extinction or large molecular column densities
\citep{LisLuc98}.  It is also the case that some very opaque lines
of sight showing CO emission are comprised entirely of diffuse material and
\HH-bearing diffuse clouds \citep{McCHin+02}:  a discussion of
such a line of sight from our own work is described in Appendix A
here.  Even in canonical dark clouds like Taurus, detailed high-resolution
mapping of the CO emission \citep{GolHey+08} reveals that much of it
originates in relatively weakly-shielded gas where \coth\ is strongly
enhanced through isotopic fractionation, implying that the
dominant form of gas phase carbon must be C\p\ \citep{WatAni+76}.

\begin{figure*}
  \psfig{figure=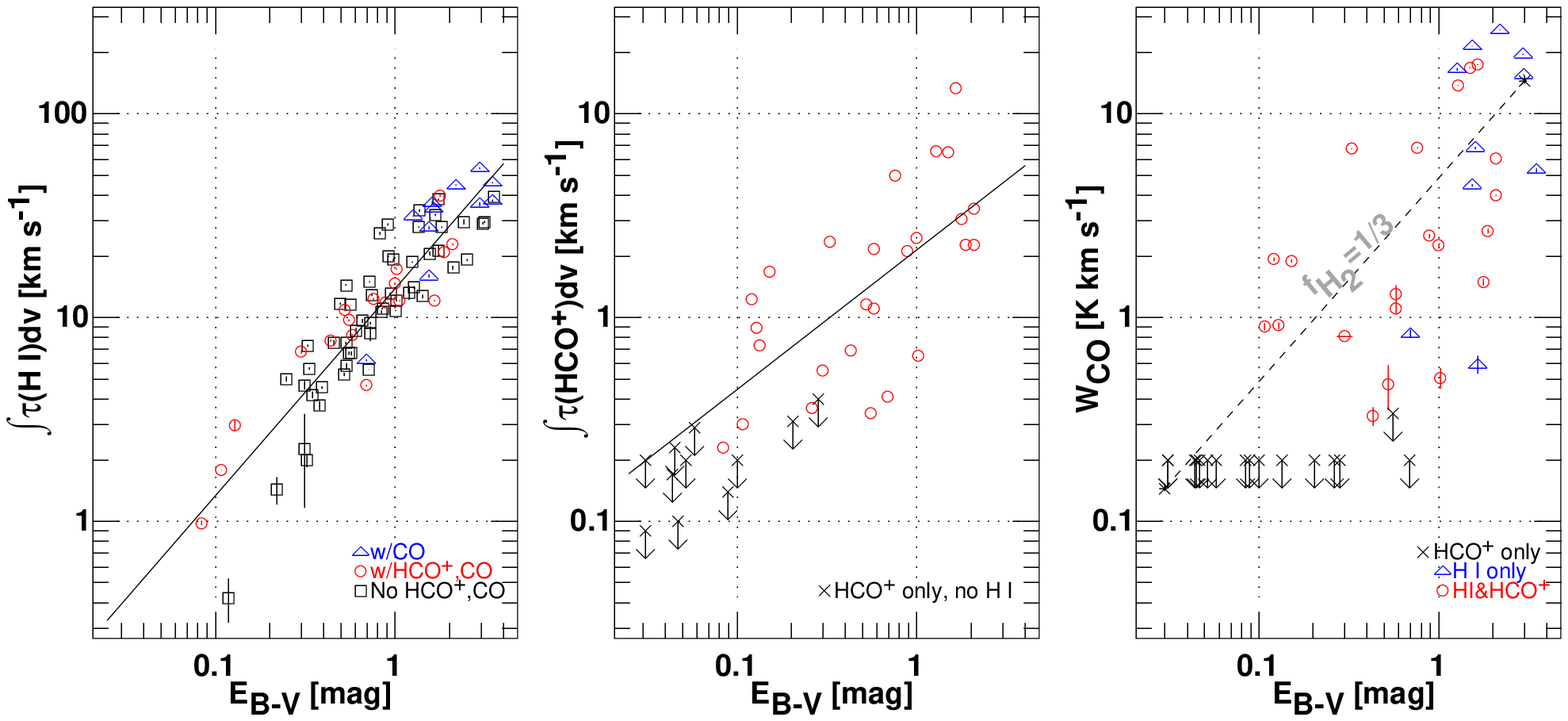,height=8cm}
  \caption[]{Atomic and molecular absorption and emission vs. total 
    reddening.  Left: Integrated VLA \HI{} optical depth from
    \cite{DicKul+83} and this work.  Middle: integrated PdBI \hcop\ optical
    depth from \cite{LucLis96} and this work.  Right: integrated ARO12m CO
    J=1-0 brightness at 1\arcmin\ resolution.  In each case the horizontal axis is the
    total line of sight reddening \EBV\ \citep{SchFin+98}. For explanation
    of the symbols used in the plots, see Sect. 3.}
\end{figure*}

Conversely, it is also the case that molecular gas is detected in the local
ISM even when CO emission is not.  Lines of sight with $\NCO \ga 10^{12}
\pcc$, $\NHH \ga 10^{19} \pcc$ have long been detectable in surveys of uv
absorption \citep{SonWel+07,BurFra+07,SheRog+07,SheRog+08}, with expected integrated
CO brightnesses as low as $\WCO = 0.001\Kkms$ \citep{Lis07}.  And, as
discussed here, mm-wave \hcop\ and CO absorption from clouds with $\NHH \ga
10^{20}~\pcc$ are also more common than CO emission along the same lines of
sight \citep[see][and Appendix A]{LucLis96,LisLuc00}.

Thus we are led to ask two questions that are of particular importance to
the use of CO emission as a molecular gas tracer.  First, where and how
does the observed local CO luminosity really originate? Second, how
completely is the molecular material represented by CO emission?  The
approach we take to address these issues is based on radiofrequency surveys
of \HI{}, \hcop\ and CO absorption and emission along lines of
sight through the Galaxy toward extragalactic background sources.  By
combining 1) measurements of extinction (constraining the total gas column
density), 2) measurements of \HI{} absorption (to determine the gas column
of atomic hydrogen), 3) the strength of \hcop\ absorption (tracing \HH\ 
directly) and 4) the integrated CO J=1-0 brightness \WCO, we relate \WCO\ 
to \NHH{} along sightlines where we have previously shown that the
intervening gas is diffuse, neither dark nor dense, and the CO column
densities are relatively small.  The results are somewhat surprising:
 although there
is much variability, the mean CO brightness per \HH-molecule \WCO/\NHH{},
i.e. the CO-\HH\ conversion factor, does not differ between
diffuse and fully molecular clouds.  Although this was phenomenologically
inferred long ago, the physical basis for it is now better understood in
terms of the radiative transfer and chemistry of \HH- and CO-bearing
diffuse and dark gas.

The plan of the present work is as follows.  Section 2 describes the
observational material that is used here, some of which is new.  Section 3
derives the CO-\HH{} conversion factor in diffuse gas.  Section 4
discusses the fraction of the local galactic CO luminosity (viewed
perpendicular to the galactic plane) that can be attributed to diffuse gas.
Section 5 discusses the physical processes at play to set the ratio of CO
brightness to \HH\ column density and explains why the same value may apply
to dark and diffuse gas. Section 6 discusses which molecular emission
diagnostics might actually be used to distinguish between the CO
contributions from diffuse and dark gas. Sections 7 and 8 present a
brief summary and discuss how our concept of the ISM might change when a
substantial portion of the observed CO emission is ascribed to diffuse
rather than dense molecular gas.

\section{Observational material}

The data used in this work are given in Tables D.1 and
D.2 of Appendix D (available online).

\subsection{\EBV}

Values of the total reddening \EBV\ along the line of sight are from the
work of \cite{SchFin+98} at a spatial resolution of 6\arcmin.  The
claimed rms error of these measurements is a percentage (16\%) of the
value.  To convert to column density we use the equivalence 
\NH = N(H I)+ 2\NHH{} $=
5.8\times 10^{21} {\rm H} ~\pcc \EBV$ established by \citet{BohSav+78}
and \citet{RacSno+09}.  Typically \AV{} = \EBV{}/3.1 \citep{Spi78}.

\subsection{\HI{} absorption}

This is mostly taken from the VLA results of \cite{DicKul+83} but a line
profile for B2251+158 (3C454.3) was made available on the website of John
Dickey and we took new \HI{} absorption profiles toward J0008+686,
J0102+584,B0528+134, B0736+017, J2007+404, J2023+318 and B2145+067 at the
VLA in 2005 May and July.

\subsection{\hcop\ absorption}

We used results from the PdBI's \hcop\ survey of \cite{LucLis96} along with
the slightly more recent results of \cite{LisLuc00} and a few additional
profiles that were taken at the PdBI in 2001-2004.

The rotational excitation of \hcop\ above the cosmic microwave background
is very weak in diffuse gas \citep{LisLuc96} so that 
$\NHCOp = 1.12\times10^{12} \pcc
\paren{\tauhcop/1\kms}$ for an assumed \hcop\ permanent dipole moment of
3.889 Debye.  This dipole moment is slightly smaller than the value used in
most of our previous work (4.07 D), increasing the inferred \hcop\ column
densities by 10\%.

\subsection{J=1-0 CO emission}

All the results quoted here are from the ARO12m antenna at 1\arcmin\ 
resolution, placed on a main-beam scale by dividing the native \Tstar\ 
values by 0.85.  Most of these profiles were used on the \Tstar\ scale in
our earlier work \citep{LisLuc96,LisLuc98,LisLuc00} but profiles toward
sources with \HI{} absorption and lacking \hcop\ absorption data (noted in
Fig.  1) and toward sources with J-names in Tables C.1 and C.2 are new.
The velocity resolution was typically 0.1 \kms\ and all spectra were taken
in frequency-switching mode and deconvolved (folded) using the EKHL
algorithm \citep{Lis97FS}.  Where upper limits on CO emission are shown, 
they are plotted symbolically at very conservative values taken over much wider 
ranges 
than are occupied by \hcop\ emission.  The contributions of such sightline
to ensemble averages of \WCO\ was taken as zero in each case.

\section{The mean $\NHH/\WCO$ ratio of diffuse gas}

\subsection{Considering whole lines of sight}

Because the target background sources are extragalactic, the lines of 
sight considered here traverse the entire galactic gas layer, crossing
the entire possible gamut  of gas phases.  However, they either have low 
extinction (at $|b| \ga 15-20$\degr) or, more often, can be decomposed 
into components whose individual molecular column densities are 
relatively small according to our previous studies of 
absorption and emission in these directions 
(see Appendix A for an example).  For instance, the highest 
CO column densities observed for individual components are
\NCO{} $\la 2\times 10^{16}\pcc$ \citep{LisLuc98}, representing
about 7\% of the free carbon column density expected for diffuse 
ISM at \AV\ = 1 mag \citep{SofLau+04}.  \coth\ is increasingly strongly
fractionated in diffuse clouds having higher \NCO{}
\citep{LisLuc98,Lis07CO}, requiring that carbon must be 
predominantly in the form of C\p.
 
\subsection{Separating the atomic and molecular gas fractions}

\begin{figure}
  \psfig{figure=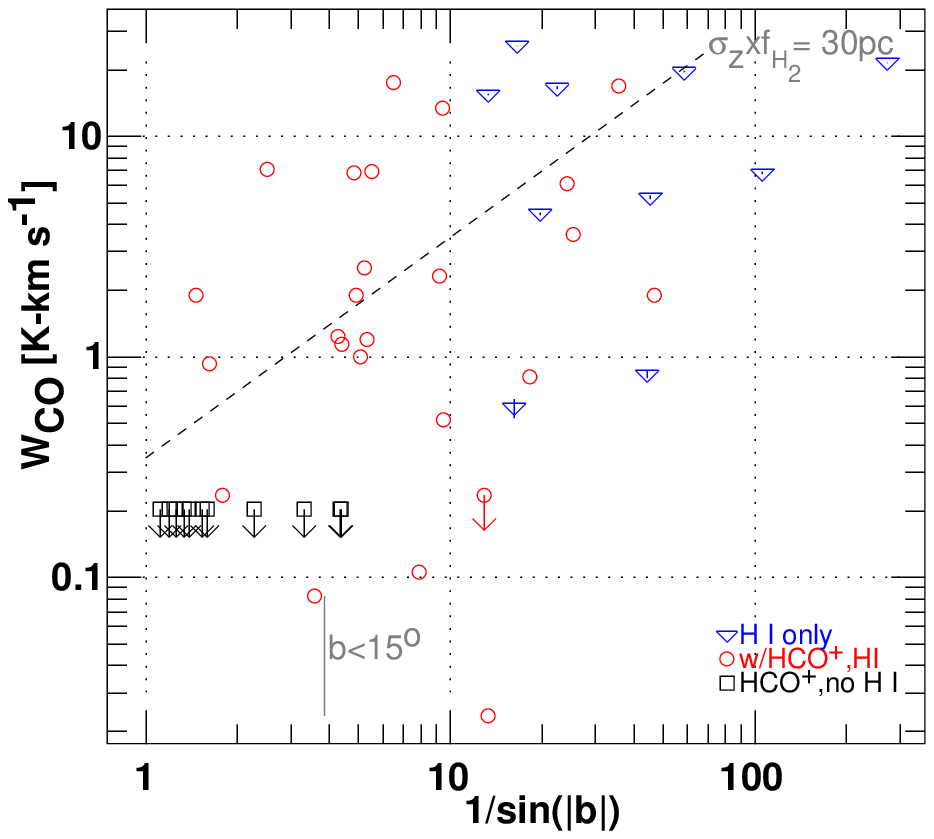,height=7.6cm}
\caption[]{Integrated CO brightness plotted against 1/sin($|b|$).
For comparison, a line is shown for the case of 
a plane-parallel Gaussian layer with vertical dispersion \sz, when 
\sz \fHH\ = 30 pc and $\NHH/\WCO = 2 \times 10^{20} ~\HH \pcc/(\Kkms)$.} 
\end{figure}

In order to derive the $\NHH/\WCO$ conversion factor, we need to estimate
\NHH{} independent of the CO emission tracer.  To do this, we could use
previous estimates of the mean fraction of \HH\ in the diffuse ISM, which
range from $\ga 25\%$ \citep{SavDra+77} in uv absorption to 40-45\% using a
chemically-based approach founded on the observed constancy of $\XCH =
\NCH/\NHH$ \citep{LisLuc02,SheRog+08,WesGal+10}.  However, as this is the
core of our argument, we take two other and more detailed approaches to
separating the atomic and molecular column densities along the actual
ensemble of lines of sight we have studied.  Both methods depend on knowing
the total column density \NH{} from the measured reddening and
both are detailed in the following subsections.


\subsection{Estimating the atomic gas fraction via
  \HI{} absorption}

In Fig. 1 at left we show the integrated \HI{} absorption plotted against
reddening.  This diagram is comprised of the entire sample of
\cite{DicKul+83} along with a handful of other sightlines observed in \HI{}
by us at the VLA and in \hcop\ at the PdBI (see Sect. 2).  Symbols
differentiate 1) those portions of the sample for which \hcop\ and CO were
observed (all sightlines observed in \hcop\ were also observed in CO
emission and most in CO absorption); 2) a few for which we only have \HI{}
absorption and CO emission data, and 3) those which lack any molecular
data.  Strictly speaking, only those lines of sight for which we have
molecular absorption line data can be proven to be composed wholly of
diffuse gas but the sample appears to be very homogeneous in terms of its
absorbing properties and many of the lines of sight lacking molecular
absorption data show CO emission well beyond the galactic extent of the
dense gas layer.

The surprisingly tight, nearly linear correlation between the integrated
\HI{} optical depth and reddening (correlation coefficient 0.90, power-law
slope 1.02) establishes the applicability of the comparison of reddening
values (which are measured on a rather coarse 6\arcmin\ spatial scale) with
\HI{} absorption measurements against the extragalactic continuum sources,
sampling sub-arcsecond beams. This excellent correlation between fan and
pencil-beam quantities testifies to the high degree to which \HI{}
absorbing gas is mixed in the interstellar gas.  The sample mean reddening
in Fig. 1 at left is $\mean{\EBV} = 1.14$ mag and the sample mean
integrated \HI{} opacity is $\mean{\tauhi} = 16.50\,\kms$ so that
 $\mean{\tauhi}/\mean{\EBV} = 1.45 $\kms/mag for the sample as a whole. 

Estimating the \HI{} column density from the \HI{} absorption must be done
with care because the atomic gas is divided between warm and cold phases
having widely differing optical depth.  Separation of the warm and cold,
absorbing and non-absorbing phases was recently considered in great detail
by \citet{HeiTro03} in a new \HI{} emission-absorption survey along many
lines of sight.  From their tabulated results, it was possible to form the
ratio of \NHI{} to \tauhi\ (a small portion of which actually arises in
warmer gas) as shown in Fig.~B.1 of the appendices and briefly
discussed in Sect. B1 there.  The sample mean ratio over all lines of sight in the
\cite{HeiTro03} survey is $\NHI /\tauhi = 2.6\pm0.2 \times 10^{20}~\pcc
/\kms$ where the error estimate (which is a range, not a standard
deviation) reflects the extent to which the ratio can be affected by sample
selection criteria based on reddening, galactic latitude, etc.  This mean
value shows very little variation when computed on sub-samples selected on
different criteria.  

It is then possible to derive the atomic gas fraction,
if we assume that our absorption sample has similar properties.. Writing
\begin{equation}
  \fHI \approx (\frac{\NHI}{\tauhi}) \times (\frac{\tauhi}{5.8\times 10^{21}\,\EBV}),
\end{equation} taking the first term from our analysis of the results of
\cite{HeiTro03} and the second from the  mean for the data shown in
Fig. 1.  The result is that $\fHI = 0.65$, so that \fHH 
 = 2 N(\HH)/N(H) = 0.35.

This estimate of the molecular gas fraction for our sample of sightlines
falls in the middle of the range of current general estimates for diffuse
gas as noted in the beginning of this Section, i.e.  $\fHH\ \ga 0.25$ from
\emph{Copernicus} corrected for sampling biases \citep{BohSav+78} and $\fHH
\approx 0.40-0.45$ from a sample of lines of sight observed toward bright
stars in optical absorption lines observed in CH \citep{LisLuc02},
given that $\XCH = \NCH/\NHH$ is nearly constant at $4.5 \times 10^{-8}$
\citep[][]{SheRog+08,WesGal+10}.

\subsection{Checking the molecular gas fraction via
  molecular chemistry}

Shown in the middle panel is the integrated \hcop\ absorption.  
As noted in Sect. 2.3 the integrated optical depth is
directly translatable into \hcop\ column density given the
  near-absence of rotational excitation in the relatively low
  density diffuse gas: $\NHCOp = 1.12 \times 10^{12} \pcc
  \paren{\tauhcop/1\kms}$.  The relative abundance of \hcop\ is known to be
nearly constant at $\XHCOp \simeq 2-3 \times 10^{-9}$ from its fixed ratio
with respect to OH in individual clouds \citep{LisLuc96,LisLuc00} and the
near-constancy of \XOH{} $\approx 10^{-7}$ \citep{WesGal+10}.

Fig. 1 shows that \hcop\ becomes readily detectable at $\EBV \ga 0.1$ mag,
which is just where \HH\ itself becomes abundant in the diffuse ISM
\citep{SavDra+77}.  When detected, \NHCOp{} shows a correlation with \EBV\ 
(correlation coefficient 0.66 and power law slope 0.7 for the points with
detected \hcop) but the larger scatter in the middle panel, compared to
that at left, suggests that the molecular portion of the gas is
less well mixed than the absorbing \HI{}.

If \XHCOp{} is assumed, a value for \fHH\ could be derived from the data in
the middle panel of Fig. 1.  Conversely, if \fHH\ = 0.35 is assumed and
sample means are used, then $\mean{\NHCOp}/(5.8\times 10^{21} \pcc
\mean{\EBV}) = 5.46\times 10^{-10}$ and $\XHCOp = \NHCOp/\NHH = 3.1 \times
10^{-9}$, consistent with the previously established value
\citep{LisLuc96,LisLuc00}.  Therefore the decomposition of the ensemble of
lines of sight appears to yield consistent results between several
independent measures of both the atomic and molecular components.

\subsection{The ensemble-averaged CO luminosity and $\NHH/\WCO$
  conversion factor}

Shown at the right in Fig. 1 is the integrated \cotw{} J=1-0 intensity
\WCO\ plotted against \EBV.  CO emission is not reliably detected except
at $\EBV > 0.3$ mag (i.e. \AV $\ga$ 1 mag).  In discussing this data, it is
important to note that values of \NCO{} have been measured in the diffuse
gas \citep{LisLuc98} and they are quite small compared to the column of
free gas phase carbon expected at \AV\ = 1 mag \citep[i.e.
$3\times10^{17}\pcc$, see][]{SofLau+04}.  Moreover, the lines of sight
having the largest values of \WCO\ are composed of several emission
components (see Appendix A for an example).  The CO emission along
these lines of sight orginates in diffuse gas where C\p\ is the dominant
form of carbon. 

If it is accepted that \fHH\ = 0.35, the  bulk CO-\HH\ 
conversion factor may be
inferred immediately from the data shown in Fig. 1.  The sample means are
\mean{\WCO} = 4.42\Kkms{} and \mean{\EBV} = 0.888 mag  or
($\mean{\NHH} = 9.01 \times 10^{20} \HH \pcc$), implying \WCO =
1\Kkms per $2.04 \times 10^{20} ~\HH \pcc$.  Rather strikingly, there is
apparently no difference in the {\it mean} CO luminosity per \HH\ in
diffuse and fully molecular gas.  For insight into the scatter
present in the ensemble of sightlines, the right-hand panel of Fig.
1 shows a line corresponding to the ensemble mean conversion factor and
$\fHH = 1/3$.   The range in \fHH\ determined for the diffuse gas,
roughly 0.25 - 0.45 or 0.35$\pm0.1$, implies a 30\% margin of
error for the method as a whole.

An alternative approach to this determination based on molecular
chemistry, comparing \WCO\ with \NHCOp as a surrogate for \NHH\
and giving similar results, is discussed in Appendix C.

\section{The proportion of CO emission arising from diffuse gas}

The similarity of the CO-\HH\ conversion factors in diffuse and fully
molecular gas must have led to confusion whereby CO emission arising in
diffuse gas has been attributed to "molecular clouds", i.e. the truism
noted in the Introduction.  To quantify this phenomenon we derive the mean
luminosity of diffuse molecular gas viewed perpendicular to the galactic
plane  $\WCO(b) \sin{\abs{b}}$ for a plane-parallel stratified gas layer
and we compare that to the equivalent luminosity perpendicular to 
the galactic plane inferred from surveys of CO emission near the galactic 
equator.

Shown in Fig. 2 is the distribution of \WCO\ with $1/\sin\abs{b}$.  For
reference a line is shown corresponding to the canonical CO-\HH\ conversion
factor and the combination $\fHH \times~ \sigma_{\rm z} = 30$ pc, in the
simplistic case that the galactic gas layer can be described by a single
Gaussian vertical component with dispersion $\sigma_{\rm z}$.  For
convenience the diffuse gas is usually described by several components 
having a range of vertical scale heights \citep{Cox05} but the neutral gas
components of the nearby ISM are not  well-described by simple
plane-parallel layers \citep[see also][]{HeiTro03} owing to local geometry
(the local bubble) combined with the scatter induced by the comparatively
long mean free paths between kinematic components.

We quote the ensemble average brightness $\mean{\WCO}$ = 4.64 \Kkms\ and
number of equivalent half-thicknesses \mean{1/\sin\abs{b}} = 19.75,
implying a mean integrated brightness 0.235\Kkms\ per galactic
half width\footnote{The actual ensemble averaged value of
  $\mean{\WCO\sin\abs{b}}$ is substantially larger $0.42\pm0.66$
  \Kkms.}. Looking down on the Milky Way vertically from afar the
integrated CO brightness of diffuse gas would be twice this, $\WCOperp =
0.47\Kkms$.

Galactic surveys of CO emission, on the other hand, calculate a mean CO
brightness per kpc of 5\Kkms/kpc at z = 0 pc in the galactic  disk
at a galactocentric distance $\Rsun = 8\,$kpc.  Note that this value
  is scaled from the result of \citet{BurGor78} which assumed $\Rsun =
  10\,$kpc.  If the molecular gas layer  sampled in these surveys 
is described by a Gaussian
vertical distribution having a dispersion \sz = 60 pc \citep{Cox05}
and z-integral $\sqrt{2\pi\sz^2} = 0.150$ kpc, the galactic survey 
result translates into an integrated 
CO brightness $\WCOperp = 5\Kkms/kpc \times 0.150$ kpc = 0.75\Kkms\ 
 when viewed vertically  across the galactic  disk
as described in Appendix D.
This is only 50\% higher than that of the diffuse CO alone.

The question then is whether the CO emission and \HH\ attributable to
diffuse gas exist in addition to that sampled in the CO surveys near the
galactic plane, or whether the galactic CO surveys incorporate a 
significant proportion of diffuse CO emission.  If the
former -- if, for instance the diffuse CO like the diffuse ISM has a larger
scale height and is a distinguishable component of the local CO emission --
the local \HH\ surface density could be higher than previously believed.
  
The total density of gas near the Sun is usually quoted as 1.2 H-nuclei
$\pccc$ from \cite{Spi78} and this is often decomposed into ``molecular''
and ``diffuse'' components with roughly 50\% attributed to each \citep[for
instance see][]{Cox05}.  The CO emissivity measured in galactic plane
surveys (5 \Kkms/kpc) conveniently converts to a local mean \HH\ density of
about $0.33\,\HH \pccc$, about half of Spitzer's total.  However, recall
that the quoted total mean density is based on the statistics of reddening
toward A-stars within a few hundred pc of the Sun \citep{Mun52} which were
very unlikely to have sampled dark cloud lines of sight.  GAIA
photometry should settle this matter, but the issue of the total
mean density of the ISM locally and relative proportions of atomic and
molecular material are not as clearly defined as is generally assumed.

\section{Rationale for a common CO-\HH conversion}

The very first discussions of the applicability of a common $\NHH/\WCO$
conversion factor \citep{Lis82,YouSco82} noted that diffuse and dense gas
at 60-100 K, or dark dense gas at 12 K, all had similar ratios \WCO/\NHH{}.
For instance $\WCO \approx 1.5\Kkms$, $\NHH{} = 5\times 10^{20}\,\kms$
toward \zoph\ (a typical diffuse line of sight) and 
$\WCO = 450\Kkms$, $\NHH{} = 2\times 10^{23}\HH\pcc$ toward Ori A. 
By comparison, a dark cloud like L204, near \zoph, with \AV\ = 5
mag has $\NHH \approx 5\times 10^{21}\pcc$, $\NCO \approx 8\times
10^{17}\pcc$ and an integrated brightness $\WCO \approx 15\Kkms$
\citep{TacAbe+00} or $\NHH/\WCO \approx 3 \times 10^{20}\HH \pcc$ (Kkms)$^{-1}$.
Comparing the two gas phases sampled in CO near \zoph\ it is apparent 
that the higher CO column density in the dark cloud is
more than compensated by the diminished brightness per CO
molecule.  The result is a nearly constant ratio of \WCO{} to \NHH{}
across phases while the brightness per CO molecule $\WCO/\NCO$ varies
widely.

The physical basis for this behavior has become more apparent recently 
with closer study of CO in diffuse gas  \citep{PetLuc+08,LisPet+09}.
To begin the discussion we rewrite the CO-\HH\ conversion factor
\NhhWco\ as
\begin{equation}
  \frac{1}{\NhhWco} = (\frac{\WCO}{\NCO})\times (\frac{\NCO}{\NHH}) = 
 (\frac{\WCO}{\NCO})\XCO, 
\end{equation}
separating the coupled and competing effects of cloud structure or
radiative transfer $(\WCO/\NCO)$ and CO chemistry (\NCO/\NHH{} = \XCO).
Simply put, the specific brightness $\WCO/\NCO$ can be shown to be higher in
warmer, subthermally-excited diffuse gas by about the same amount (a
factor 30-50) that \XCO{} is lower: $\mean{\XCO{}} = 3\times 10^{-6}$ for
the diffuse gas \citep{BurFra+07} compared to $\approx 10^{-4}$ in dark gas
where the carbon is very nearly all in CO.

As noted by \cite{GolKwa74} in the original exposition of the LVG model for
radiative transfer, \WCO/\NCO{} will be much greater when the excitation of CO
is weak -- when the kinetic temperature is much greater than the J=1-0
excitation temperature.  Moreover when CO is excited somewhat above 
the cosmic microwave background but well below the kinetic temperature, 
the brightness of the CO J=1-0 line will be linearly proportional to \NCO{} 
even when the line is quite optically thick \citep[again, see][]{GolKwa74}.  
As M. Guelin
pointedly reminded us, this occurs because weak excitation means that
there is also little collisional de-excitation so that the gas merely
scatters emitted photons until they eventually escape.  As \cite{GolKwa74}
showed, this proportionality between brightness and column density
persists until the opacity is so very large that the transition approaches
thermalization through radiative trapping.

\newcommand{\pccpKkms}[1]{\emr{\,#1\,cm^{-2}/(K\,km\,s^{-1})}}

The discussion of the previous paragraph also applies to other molecules, 
but because CO has such a small dipole moment the proportionality 
between CO brightness and column density is only weakly dependent on 
ambient physical conditions:  a
nearly universal ratio $\NCO/\WCO = 10^{15} \pccpKkms{CO}$ can be
calculated for diffuse gas using recent excitation
cross-sections \citep{Lis07}.  This is in excellent agreement with measured
values of \NCO{} and CO J=1-0 excitation temperatures in the diffuse gas
seen toward stars in uv absorption \citep{SonWel+07,BurFra+07,SheRog+08} or
at mm-wavelengths in absorption against distant quasars
\citep{LisLuc98}.  For the observed value $\mean{\XCO{}} = 3\times 10^{-6}$
\citep{BurFra+07} the \NHH/\WCO\ conversion ratio in diffuse clouds 
is \NHH{}/\WCO $= 10^{15}/3\times 10^{-6} = 3.3 \times 10^{20}\pccpKkms{\HH}$.

Finally, note that even if the ratio \WCO/\NCO{} is not constant between 
gas phases, it is still the case that $ \WCO \propto \NCO$ separately 
in either the dense or diffuse gas.  For the diffuse gas the 
proportionality is based in the microphysics of CO radiative transfer 
a la \cite{GolKwa74}.  For the dark cloud case, note that there is a 
fixed ratio of \NCO{}/\NHH{} when the the gas-phase carbon is in CO 
and the hydrogen is in \HH\ so that a \WCO-\NHH{} conversion 
is fully equivalent to a \WCO-\NCO{} conversion.



\section{Discriminating between emission from diffuse and dense gas}

There are ways in which mm-wave molecular emission differs between dense
and diffuse gas, even if not in \cotw.  Emission from molecules like CS,
HCN and \hcop\ having higher dipole moments is generally thought to single
out denser gas than does CO, especially in extreme environments
\citep{WuEva+05}.  Note, however, that surveys of the Milky Way
galactic plane find widely-distributed emission in 
\hcop, CS, HCN, etc.  with intensity ratios of 1-2\% relative 
to \WCO\ from essentially all features detected in CO \citep{Lis95,HelBli97}.

Relatively little is known of the emission from mm-wave species in diffuse
gas beyond that from CO.  Most common is emission from \hcop\ because it is
chemically ubiquitous and somewhat more easily excited owing to
its positive charge and high dipole moment.  Although \hcop\ emission is
weak in the example shown here in Appendix A it appears at levels
$\ga 1$\% of \WCO\ in portions of the diffuse cloud around \zoph\ 
or in the Polaris flare \citep{LisLuc94,Lis97,FalPin+06}.  Therefore 
\hcop\ emission is probably not
a good discriminator but CS and HCN appear with high abundance only when
\NHCOp{} $\ga 10^{12}\pcc$ or $\NHH > 5\times 10^{20} \pcc$ and should be
much more weakly excited in low density gas.  In any case, searching for
emission that is 100 times weaker than \WCO\ may not be an effective use of
observing time and only in very dense, warmer gas like that found in
massive, OB star-forming regions like Ori A are the higher dipole moment
molecules substantially brighter than 1-2\% relative to \WCO.

A more effective method of discriminating between CO emission from diffuse
and dark or dense gas is afforded by \coth.  Although the abundance of
\coth\ is enhanced by fractionation (see the example in Appendix A)
lowering the observed \cotw/\coth\ brightness temperature ratios
\citep{LisLuc98,Lis07,GolHey+08}, those ratios are still noticeably higher
in diffuse gas.  Typically they are $\ga 10-15$ instead of $\la 3-5$ as
seen in dark clouds or in surveys of the inner-Galaxy gas in the galactic plane
\citep{BurGor78}.  Recall that the mean \cotw/\coth\ brightness ratio
nearly doubles across the galactic disk \citep{LisBur+81}, which was
another, earlier indication that molecular gas near the Solar Circle has a
high proportion of diffuse material.

To summarize, we suggest that the most efficient way to ascertain the
origin of CO emission is to compare \cotw\ and \coth\ brightnesses because
emission from \coth\ is much stronger than emission from \hcop, CS, HCN
etc, and because there is actually less ambiguity in the brightness ratios
relative to \WCO.

\section{Summary}

In Sect. 2 and 3 we described and considered a sample of lines of sight
studied in \HI{} and molecular absorption and known to be comprised
of diffuse gas.  Their molecular component shows features whose CO,
\hcop\ and other molecular column densities are small compared to those of
dark clouds (in the case of CO, at least 30 times smaller).   There is 
often quite substantial fractionation of \coth\ (indicating that the dominant 
form of carbon is C\p) and the rotational excitation of CO is sub-thermal
with implied cloud  temperatures typical of those determined directly 
for diffuse \HH\ in optical/uv surveys, i.e. 30 K or more.
  
Using an externally-determined value for the ratio of total \HI{} column
density to integrated \HI{} absorption and the standard equivalence between
reddening and \NH{} we derived the molecular gas fraction for this sample
to be \fHH\ = 0.35, in the middle of the range of other estimates for the
diffuse ISM as a whole based on optical (mainly CH) and uv (\HI{} and \HH)
absorption studies.

We showed that this estimate for \fHH\ implies the same value $\XHCOp =
3\times 10^{-9}$ that was previously determined from comparisons of OH and
\hcop\ column densities in individual clouds.  We then compared measured CO
brightnesses with the inferred molecular gas column densities to
derive the ensemble mean $\NHH/\WCO$ conversion factor.  Surprisingly, We
found this mean to be just equal to the locally-accepted value $2.0\times
10^{20} \HH$ /(\Kkms) for ``molecular'' gas believed to reside in dense
dark fully-molecular clouds near the galactic equator.

Such exact agreement is probably something of an accident of sampling, but
the fact that diffuse and dark gas would have very similar $\NHH/\WCO$
conversion factors, which had been inferred empirically long ago, now has a
firmer physical basis.  In Sect. 5 we explained it as the result of the
brightening of CO J=1-0 emission per CO molecule that was theoretically
predicted for warmer more diffuse gas by \citet{GolKwa74}, which
compensates for the lower relative abundance \XCO{} there.  The mean
CO abundance observed in optical absorption in diffuse clouds 
$\mean{\XCO} = 3\times 10^{-6}$, combined with the observed and expected
brightness per CO molecule, \WCO/\NCO{} = 1 \Kkms/$10^{15} {\rm CO}\pcc$,
can be be combined to form an CO-\HH\ conversion factor of 
\NHH{}/\WCO $= 10^{15}/3\times 10^{-6} = 3.3 \times 10^{20}\pccpKkms{\HH}$.

In Sect. 4 we derived the expected brightness of diffuse gas viewed
perpendicular to the galactic plane from afar, 0.47\Kkms, and compared that
to the value expected from surveys of CO emission in the galactic plane,
combined with a narrow (60 pc dispersion) Gaussian vertical distribution;
that is 0.75\Kkms.  This suggests that there has been confusion in the
general attribution of CO emission to ``molecular clouds'' when in fact
much of it arises in the diffuse ISM.  This view is consistent with the
motivations discussed in the Introduction, whereby CO emission is
increasingly being found along lines of sight lacking high extinction and
whereby CO emission seen along dark lines of sight are found (through
molecular absorption studies and in other ways) to originate in
components having the relatively small molecular number and column
densities typical of diffuse gas.  An example of such a line of sight is
given in Appendix A here.

We briefly discussed in Sect. 4 the decomposition of the total mean density
of neutral gas in the nearby ISM, 1.2 H $\pccc$ \citep{Spi78}, into its
atomic and molecular constituents.  We noted that although the balance is
generally believed to be roughly 50-50 \citep{Cox05}, some emission might
shift to the diffuse side of the balance sheet if CO emission is
reinterpreted.  Moreover, we pointed out that the molecular contribution
to the true local mean density from large-scale galactic CO surveys in 
the galactic plane should be questioned more generally because it is unclear 
to what extent Spitzer's estimate, based on the earlier optical work of 
M\"unch, incorporates the contribution of optically-opaque gas.
 
Although the ability to discriminate between the separate contributions to
\WCO\ from diffuse and darker, denser gas is limited when only \cotw\ is considered,
it should be possible to infer the nature of the host gas using other
emission diagnostics (see Sect. 6).  The most efficient of these is
probably the brightness of \coth, which, although enhanced by
fractionation, is still substantially weaker, relative to \WCO, in diffuse
gas.  Searching for emission from species having higher dipole moments such
as CS J=2-1 and HCN (and probably not \hcop\ because it is chemically so
ubiquitous and more easily excited) are alternatives that might require
somewhat longer integration times.

\section{Discussion: Interpreting a sky occupied by CO emission from diffuse gas}

The usual interpretation of CO sky maps, galactic surveys, etc, is that CO
emission mostly traces dark and or "giant" molecular clouds (GMC) composed
of dense cold gas occupying a very small fraction of the interstellar
volume at high thermal pressure within an ISM that may confine them via its
ram or turbulent pressure if they are not gravitationally bound.  The
balance between GMC and diffuse atomic material may be controlled by
quasi-equilibrium between local dynamics and the overlying weight of the
gas layer but the molecular material inferred from CO emission is generally
believed to be that which is most nearly on the verge of forming stars, for
instance through the Schmidt-Kenicutt power-law relation between star
formation rate and gas surface density\footnote{It is also recognized that
  more precise tracers of the high-density star-forming material may be
  needed in extreme environments such as ULIRG \citep{WuEva+05}.}
\citep{LerWal+08,BigLer+08}.

By contrast, CO emission from diffuse molecular gas originates within a
warmer, lower-pressure medium that occupies a much larger fraction of the
volume and contributes more substantially to mid-IR dust or PAH emission
but only has the requisite density and chemistry to produce CO molecules
and CO emission (since $\WCO \propto\NCO$) over a very limited portion of
that volume.  In this case a map of CO emission is a map of CO abundance
and CO chemistry first, and only secondarily a map of the mass even if the
mean CO-\HH\ conversion ratio is (as we have shown) "standard".  Moreover,
although CO emission traces the molecular column density \NHH{} quite
decently where \WCO\ is at detectable levels, it arises in regions that are
not gravitationally bound or about to form stars.  The CO sky is
mostly an image of the CO chemistry.
 
\begin{acknowledgements}
  The National Radio Astronomy Observatory is operated by Associated
  Universites, Inc. under a cooperative agreement with the US National
  Science Foundation.  The Kitt Peak 12-m millimetre wave telescope is
  operated by the Arizona Radio Observatory (ARO), Steward Observatory,
  University of Arizona.  IRAM is operated by CNRS (France), the MPG
  (Germany) and the IGN (Spain). This work has been partially funded by the
  grant ANR-09-BLAN-0231-01 from the French {\it Agence Nationale de la
    Recherche} as part of the SCHISM project.  We thank Bob Garwood for
  providing the H I profiles of \cite{DicKul+83} in digital form.
\end{acknowledgements}

\bibliographystyle{apj}


\begin{appendix}

\section{NRAO150: An example of a dark line of sight comprised of diffuse gas}

\begin{figure}
  \psfig{figure=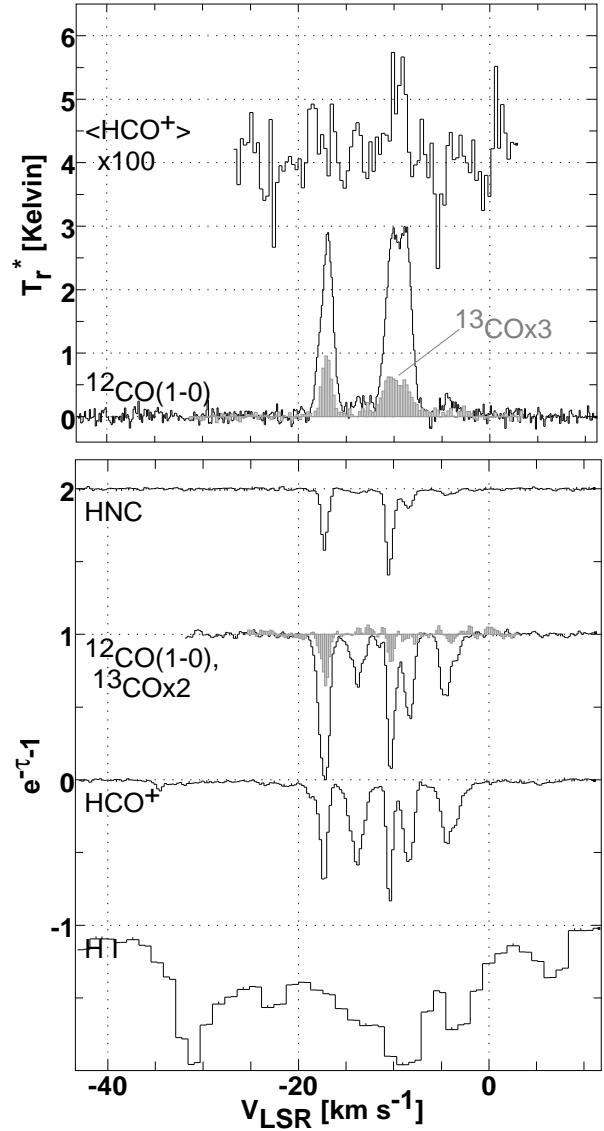,height=15cm}
\caption[]{Line profiles toward and near B0355+508 = NRAO150. 
Bottom: absorption line profiles
  of H I, \hcop, \cotw, \coth{} (multiplied by 2) and HNC; \HI{}
  absorption and emission are present over a much broader velocity range
  than shown here.  Top: Emission from \cotw, \coth{} (scaled upward by a
  factor 3) and \hcop\ (scaled upward by a factor 100).  The \hcop\ profile
  is an average over a 3.5\arcmin\ region around the continuum (to avoid
  absorption).  See Appendix A.}
\end{figure}

The estimated total extinction along this comparatively low-latitude line
of sight at l=150.4\degr, b=-1.6\degr\ (see Table D.2) is \EBV\ = 1.5 mag
or \AV $\approx 5$ mag but it would be quite opaque even if only the atomic
gas were present.  A lower limit on \NHI{} from the integrated 21 cm
emission of the nearest profile in the Leiden-Dwingeloo Survey
\citep{HarBur97} in the optically thin limit is $\NHI \ga 7.4
\times 10^{21} \pcc$, implying \EBV\ $\ga 1.27$ mag.  The H I column
density derived by taking the ratio of \NHI{} to \HI{} absorption
as discussed in Sect. 3 here is, understandably, slightly larger,
$\NHI = 1.1\times10^{22}\pcc$.

We show in Fig. A.1 various absorption and emission profiles along
and around the line of sight to NRAO150 aka B0355+508.  We have
published various analyses of this line of sight in the references noted
below, and most recently we synthesized the CO emission in a 90\arcsec\
region around NRAO150 at 6\arcsec\ resolution \citep{PetLuc+08}.
\HI{} absorption and emission extend well outside the narrow
kinematic interval shown here.  The weak \hcop\ absorption at -35 \kms\ is
real, as is the broad wing extending up to -25 \kms.

CO emission is fairly strong in this direction, \WCO\ = 17 K \kms,
nominally implying 2\NHH{} $\approx 7 \times 10^{21}$, comparable to
\NHI{}, but molecular absorption spectra of \hcop\ and CO are much
richer than the CO emission.  The \hcop\ absorption spectrum
\citep{LucLis96,LisLuc00} shows five prominent components each having
\NHCOp{} $\approx 1.3 \times 10^{12}\pcc$ \citep{LucLis96} or 2\NHH{}
$\approx 9\times10^{20} \pcc$ implying \EBV\ = 0.15 mag per component
associated with \HH\ if $\XHCOp = 3 \times 10^{-9}$ as discussed in Sect.
3.  The \HH, \hcop\ and OH column densities of these components are each
nearly equal to what is seen locally along the line of sight to \zoph\ at
\AV\ = 1 mag \citep{Mor75,VanBla86,Lis97}.

Further evidence of the diffuse nature of the gas is given by the
fractionation of \coth\ in CO; $\N{\cotw}/\N{\coth} = 15\pm 2$,
  $25\pm4$ and $32\pm 13$ in the components at -17, -11 and -4 \kms,
respectively and $\N{\coth}/\N{\coei} > 36$, $> 54$ and $> 25$ at
the $2\sigma$ level in these components \citep{LisLuc98}.

In emission, the \cotw/\coth\ brightness ratios are 12 and 30 for the two
strong kinematic components, reflecting both the fractionation and the fact
that $\WCO \propto \NCO$ in the diffuse gas regime as discussed in the text
here.

\hcop\ emission is weak in Fig. A.1.  The profile shown
\citep[from][]{LucLis96} is an average of positions around the continuum
source to avoid contamination from absorption.  The low levels of \hcop\ 
emission seen toward our sample of background continuum sources can be
understood as arising from relatively low density gas ($n_\HH \la 100
\pccc$) when the electron fraction is as high as expected for diffuse gas,
i.e.  $2\times10^{-4}$ \citep{LucLis94,LucLis96}.

\section{The ratio of total to absorbing \HI{}}

Shown in Fig. B.1 is a plot of the data from the tables of \cite{HeiTro03}
that were used in Sect. 3 to convert the \tauhi\ measurements in Fig. 1 to
a total quantity of \HI{}.  The plot shows a regression line (power-law
slope 0.84) fit to data points with \EBV\ $> 0.09$ mag (the range occupied
by the \hcop\ {\it detections} in Fig. 1) to point out a slight upturn at
low \tauhi.  The sample means are largely unaffected by setting various
sample selection criteria.

\begin{figure}
  \psfig{figure=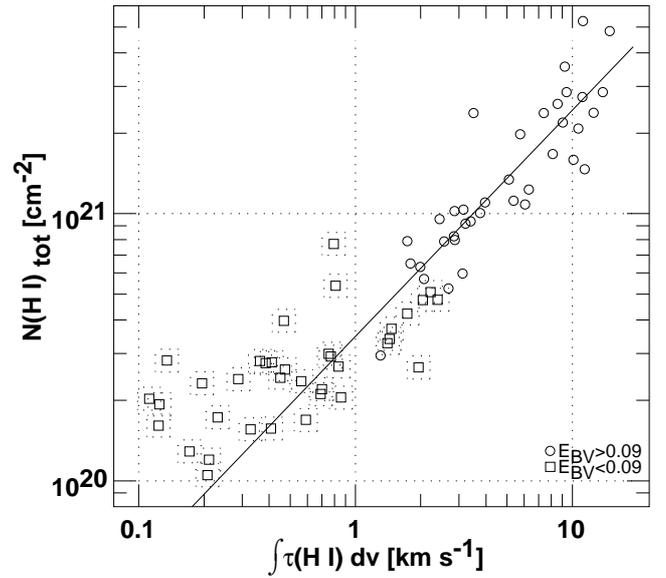,height=7.6cm}
\caption[]{Total hydrogen column density vs. integrated \HI{} optical depth 
  for the sources studied by \cite{HeiTro03}.  Lines of sight with \EBV\ 
  $<$ 0.09 mag were not included in the regression fit, to point out the
  upturn at low \tauhi. See Appendix B.}
\end{figure}

\section{A chemistry-based determination of \NHH/\WCO }

\begin{figure}
  \psfig{figure=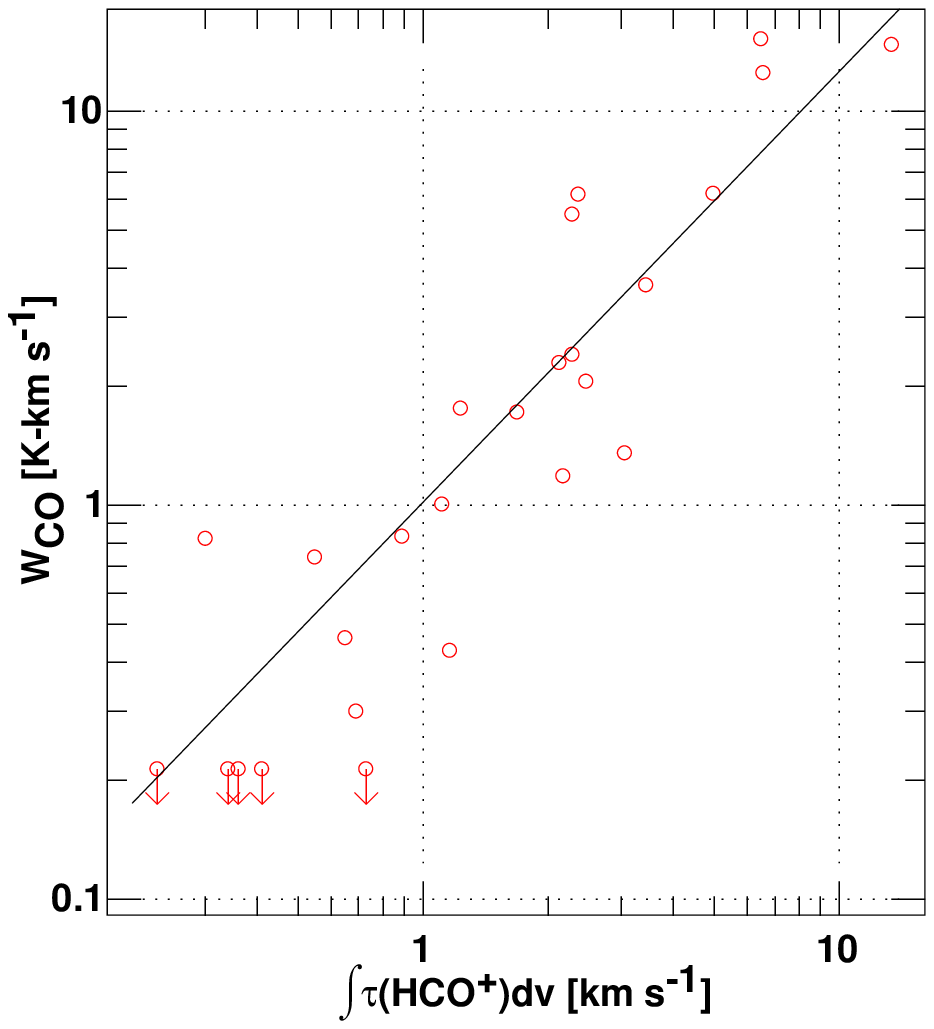,height=9cm}
\caption[]{Integrated CO J=1-0 brightness plotted against the integrated
  \hcop\ J=1-0 optical depth.  $\NHCOp = 1.12\times10^{12} \pcc
  \paren{\tauhcop/1\kms}$.  See Appendix D.}
\end{figure}

It is also possible to determine \WCO/\NHH{} without the H I measure
or formally estimating \fHH, although we preferred not to do this
in the main discussion.  In Fig. C.1 we show the variation of 
\WCO\ with \tauhcop.
CO appears reliably at detectable levels $\WCO \ga 0.3\Kkms$, $\NCO \ga
3x10^{14} \pcc$ when $\NHCOp \ga 3\times 10^{11} \pcc$ or $\NHH \ga
\NHCOp{}/3\times 10^{-9} = 10^{20}~\pcc$.  If $\XHCOp{} = 3\times 10^{-9}$
the ensemble mean values $\mean{\WCO} = 3.45\Kkms$, $\mean{\tauhcop} =
2.38\,\kms$ imply $\WCO = 1\Kkms$ per $2.6\times\ 10^{20}~\HH\pcc$, just 30\%
above that derived in Sect. 3.5.

  The near linearity of the \NCO{}-\NHCOp{} relationship in Fig. C.1
  results from bulk averaging over whole lines of sight: given 
  the same general mix of conditions, an ensemble of richer and poorer
  or shorter and longer sightlines will show proportionalities between 
  almost any two quantities in this way.  As shown in Fig. A.1
  there is no such proportionality on a per-component basis. 
  In detail, and with much scatter, the overall chemical variation is 
  approximately $\NCO \propto (\NHH)^2$  \citep{Lis07,SheRog+08}. 

\section{Calculating the CO brightness from galactic survey results}

The statistics of observing the clumpy galactic molecular cloud distribution 
are Poisson \citep{BurGor76,BurGor78} so the integrated CO brightness 
\WCO(r) accumulated when traversing a path of length r in the galactic plane is

\begin{equation}
  \WCO(r) = \ZWCO (1-\exp{(-r/\Lambda)})
\end{equation}

where \ZWCO{} is the characteristic brightness of a clump (GMC) and
$\Lambda$ is the geometric mean free path between clumps.  Although
it is possible to derive \ZWCO{} and $\lambda$ separately, galactic
survey results are given in terms of a hybrid quantity 
A$_{\rm CO}$ whose units are \Kkms{} per kpc corresponding to evaluating 
\WCO(r) when r $<< \lambda$, i.e. \WCO(r) = (\ZWCO/$\lambda$) r 
= A$_{\rm CO}$ r.   The coefficient A$_{\rm CO}$ is closely related
to the mean density: just convert \WCO\ to \NHH{}. For H I the
integrated brightness per unit distance is directly converted into
a mean density n(H I), if it is assumed that the gas is optically
thin.

The brightness of the CO cloud ensemble viewed vertically through
the galactic disk is then just  A$_{\rm CO}\,\Delta$z, where $\Delta$z
is the equivalent thickness of the disk.  For a Gaussian
vertical distribution with dispersion \sz, 
$\Delta {\rm z}  = (2\pi)^{1/2} \sz$.

\Online{}

\section{Data}

The data shown in Figs. 1  are tabulated in Tables D.1 and D.2.  
The sources of these data are discussed in Sect. 2.

\begin{table*}
\caption[]{Data used in this work}
{
\begin{tabular}{lcccccc}
\hline
Source &    l &   b    & \EBV$^a$ & \tauhi$^b$ & \tauhcop$^c$ & \WCO$^d$ \\
       & \degr & \degr & mag      & \kms       & \kms &      \Kkms \\
\hline
   B1748-253& 3.745& 0.635& 7.86& 45.37(0.40)& &  \\
   B2005+403& 6.816& 4.302& 0.69& 4.67(0.05)& 0.41(0.02)& $<$0.20 \\
   B1730-130& 12.032& 10.812& 0.53& 10.91(0.17)& 1.16(0.02)& 0.47(0.12) \\
   B1908-210& 16.857& -13.219& 0.28& & $<$0.30& $<$0.20 \\
   B1819-131& 17.910& 0.372& 7.99& 63.71(1.00)& &  \\
   B1817-098& 20.711& 2.293& 1.55& 20.62(0.49)& &  \\
   B1819-096& 21.047& 1.957& 3.08& 29.01(0.40)& &  \\
   B1829-106& 21.347& -0.629& 11.56& 66.05(0.53)& &  \\
   B1741-038& 21.591& 13.128& 0.58& 8.20(0.11)& 1.11(0.10)& 1.11(0.07) \\
   B1849+005& 33.498& 0.194& 16.93& 117.62(0.87)& &  \\
   B1749+096& 34.920& 17.644& 0.09& & $<$0.14& $<$0.20 \\
   B1909+049& 39.694& -2.244& 2.51& 19.23(0.22)& &  \\
   B1910+052& 40.100& -2.336& 2.10& 17.68(0.41)& &  \\
   B1843+098& 41.112& 5.772& 0.57& 6.71(0.16)& &  \\
   B1915+062& 41.605& -2.928& 1.43& 12.78(0.27)& &  \\
   B1909+161& 49.658& 2.907& 1.54& 15.98(0.33)& & 4.47(0.06) \\
   B1905+190& 52.496& 5.591& 0.66& 9.69(0.23)& &  \\
   B1923+210& 55.557& 2.264& 1.87& 21.01(0.28)& 2.28(0.12)& 2.66(0.07) \\
   B1950+253& 62.366& -0.956& 3.15& 29.38(0.33)& &  \\
   B1901+319& 63.029& 11.757& 0.12& 0.42(0.10)& &  \\
   B1641+399& 63.455& 40.948& 0.04& & $<$0.09& $<$0.20 \\
   B2145+067& 63.656& -34.072& 0.08& 0.98(0.02)& 0.23(0.07)& $<$0.20 \\
   B2007+249& 64.048& -4.334& 1.27& 14.12(0.21)& & 16.62(0.25) \\
   J2023+319& 71.397& -3.093& 1.06& 12.11(0.08)& 1.55(0.04)&  \\
   B2015+33A& 72.226& -0.978& 3.48& 46.23(0.34)& &  \\
   B2015+33B& 72.226& -0.981& 3.48& 37.55(0.43)& & 5.31(0.11) \\
   B2023+336& 73.129& -2.368& 2.08& 22.90(0.26)& 3.43(0.02)& 6.04(0.04) \\
   B2048+313& 74.585& -8.045& 0.22& 1.43(0.22)& &  \\
   B2013+370& 74.866& 1.224& 1.78& 39.67(0.60)& 3.12(0.02)& 1.49(0.06) \\
   B1954+513& 85.298& 11.757& 0.15& & 1.68(0.06)& 1.89(0.04) \\
   B1823+568& 85.739& 26.080& 0.06& & $<$0.20& $<$0.20 \\
   B2251+158& 86.111& -38.184& 0.11& 1.78(0.01)& 0.30(0.01)& 0.91(0.04) \\
   B2037+511& 88.808& 6.040& 1.02& 17.32(0.13)& 0.65(0.17)& 0.51(0.06) \\
   B2022+542& 90.093& 9.665& 0.35& 4.17(0.16)& &  \\
   B2055+508& 90.378& 3.533& 2.95& 36.22(0.83)& & 19.65(0.12) \\
   B2106+494& 90.528& 1.305& 2.97& 54.15(0.52)& & 15.53(0.09) \\
   B2030+547& 91.129& 8.988& 0.38& 3.73(0.20)& &  \\
   B2200+420& 92.590& -10.441& 0.33& 3.74(0.04)& 2.36(0.03)& 6.78(0.05) \\
   B2154+483& 95.584& -4.860& 0.53& 7.51(0.13)& &  \\
   B2111+620& 100.287& 9.429& 0.84& 10.72(0.18)& &  \\
   B2146+608& 102.570& 5.713& 0.85& 11.04(0.11)& &  \\
   B2201+62S& 104.940& 5.833& 0.73& 9.44(0.14)& &  \\
   B1928+738& 105.625& 23.541& 0.13& & 0.73(0.03)& $<$0.20 \\
   B2341+535& 112.952& -7.745& 0.33& 5.62(0.09)& &  \\
   B2255+702& 113.596& 9.707& 0.52& 5.27(0.10)& &  \\
   B2357+554& 115.718& -6.503& 0.31& 4.65(0.23)& &  \\
   B2357+55B& 115.719& -6.498& 0.31& 2.27(1.10)& &  \\
   B2348+644& 116.513& 2.555& 1.26& 31.56(0.17)& &  \\
   B0012+610& 118.548& -1.264& 1.65& 34.20(0.15)& & 0.59(0.06) \\
   B0016+731& 120.644& 10.728& 0.32& 1.99(0.14)& &  \\
   B0041+660& 122.253& 3.449& 2.18& 44.63(0.32)& & 25.80(0.11) \\
   B0052+681& 123.351& 5.503& 1.00& 10.79(0.06)& &  \\
\hline
\end{tabular}}
\\
$^a$ Reddening from \cite{SchFin+98} rms error is 16\% \\
$^b$ \tauhi\ from \cite{GarDic89} and (for J-sources) this work see Sect. 2 \\
$^c$ \tauhcop\ from \cite{LucLis96}, \cite{LisLuc00} and this work see Sect. 2 \\
$^d$ \WCO\ from \cite{LucLis96}, \cite{LisLuc98} and this work see Sect. 2 \\
\end{table*}

\begin{table*}
\caption[]{Data used in this work (continued)}
{
\begin{tabular}{lcccccc}
\hline
Source &    l &   b    & \EBV$^a$ & \tauhi$^b$ & \tauhcop$^c$ & \WCO$^d$ \\
       & \degr & \degr & mag      & \kms       & \kms &      \Kkms \\
\hline
   B0056+666& 123.782& 3.992& 1.20& 13.30(1.05)& &  \\
   J0102+584& 124.426& -4.436& 0.56& 9.75(0.10)& 0.34(0.01)& $<$0.34 \\
   B0107+562& 125.637& -6.231& 0.39& 4.56(0.07)& &  \\
   B0125+628& 127.109& 0.538& 1.60& 36.54(0.59)& & 6.81(0.09) \\
   B0212+735& 128.927& 11.964& 0.76& 12.36(0.13)& 4.98(0.20)& 6.81(0.06) \\
   B0205+614& 132.064& 0.210& 1.55& 27.59(0.45)& & 21.54(0.06) \\
   B0224+671& 132.122& 6.234& 1.00& 14.73(0.24)& 2.46(0.07)& 2.27(0.06) \\
   B0241+623& 135.636& 2.431& 0.73& 8.38(0.71)& &  \\
   B0323+55A& 143.890& -1.057& 1.75& 37.96(0.30)& &  \\
   B0300+471& 144.986& -9.863& 0.25& 4.99(0.14)& &  \\
   B0954+658& 145.746& 43.132& 0.12& & 1.23(0.29)& 1.94(0.04) \\
   B0332+534& 145.952& -1.681& 1.68& 31.89(0.35)& &  \\
   B0334+506& 147.809& -3.895& 1.24& 18.77(0.14)& &  \\
   B0430+587& 148.581& 7.536& 0.54& 5.80(0.19)& &  \\
   B0355+508& 150.377& -1.604& 1.50& 42.79(1.10)& 6.48(0.03)& 16.82(0.12) \\
        3C84& 150.577& -13.261& 0.05& & $<$0.10& $<$0.20 \\
   B0442+506& 155.877& 3.460& 0.94& 13.14(0.21)& &  \\
   B0435+487& 156.417& 1.317& 1.35& 27.84(0.22)& &  \\
   B0235+164& 156.772& -39.108& 0.03& & $<$0.20& $<$0.20 \\
   B0404+429& 156.780& -6.586& 0.49& 11.73(0.13)& &  \\
   B0420+417& 159.705& -5.382& 0.74& 12.93(0.12)& &  \\
   B0458+477& 159.712& 3.653& 0.72& 15.07(0.14)& &  \\
   B0406+386& 159.845& -9.484& 1.02& 12.12(0.12)& &  \\
   B0429+416& 160.965& -4.342& 0.56& 11.55(0.04)& &  \\
   B0415+379& 161.686& -8.788& 1.66& 12.14(0.14)& 13.34(0.65)& 17.45(0.11) \\
   B0442+39A& 164.109& -3.656& 0.56& 6.69(0.12)& &  \\
   B0509+406& 166.502& 0.916& 0.53& 14.45(0.16)& &  \\
   B0552+398& 171.647& 7.285& 0.44& 7.71(0.12)& 0.69(0.08)& 0.33(0.04) \\
   B0923+392& 183.709& 46.165& 0.10& & $<$0.09& $<$0.20 \\
   B0601+204& 189.566& -0.640& 1.37& 33.58(0.56)& &  \\
   B0528+134& 191.368& -11.012& 0.89& 11.88(0.20)& 2.14(0.02)& 2.53(0.06) \\
   B0629+160& 196.582& 3.204& 0.46& 7.54(0.25)& &  \\
   B0622+147& 196.983& 1.103& 0.82& 26.07(0.46)& &  \\
   B0629+104& 201.531& 0.508& 1.81& 27.96(0.25)& &  \\
   B0630+082& 203.544& -0.272& 0.91& 28.65(0.33)& &  \\
   B0624-058& 215.439& -8.067& 0.71& 5.55(0.02)& &  \\
   J0008+686& 215.752& -13.253& 1.28& & 7.16(0.75)& 13.79(0.05) \\
   B0605-085& 215.752& -13.523& 0.58& & 2.17(0.25)& 1.31(0.13) \\
   B0736+017& 216.990& 11.380& 0.13& 2.96(0.17)& 0.89(0.10)& 0.92(0.04) \\
   B0607-157& 222.611& -16.183& 0.26& & 0.36(0.09)& $<$0.20 \\
   B0727-115& 227.768& 3.140& 0.30& 6.81(0.14)& 0.55(0.02)& 0.81(0.04) \\
   B0733-174& 233.585& 1.444& 0.92& 20.00(0.13)& &  \\
   B0709-206& 233.670& -5.021& 0.98& 19.22(0.28)& &  \\
   B0704-231& 235.337& -7.218& 0.33& 7.28(0.10)& &  \\
   B1055+018& 251.513& 52.775& 0.20& & $<$0.24& $<$0.20 \\
       3C273& 289.954& 64.360& 0.03& & $<$0.08& $<$0.20 \\
       3C279& 305.107& 57.062& 0.05& & $<$0.07& $<$0.20 \\
   B1334-127& 320.026& 48.374& 0.04& & $<$0.17& $<$0.20 \\
   B1714-397& 347.748& -1.142& 3.53& 39.45(0.20)& &  \\
   B1705-353& 350.339& 2.768& 1.74& 21.39(1.30)& &  \\
   B1714-336& 352.735& 2.393& 2.42& 29.58(0.88)& &  \\
   B1711-285& 356.516& 5.884& 0.60& 8.64(0.27)& &  \\
\hline
\end{tabular}}
\\
$^a$ Reddening from \cite{SchFin+98} rms error is 16\% \\
$^b$ \tauhi\ from \cite{GarDic89} and (for J-sources) this work see Sect. 2 \\
$^c$ \tauhcop\ from \cite{LucLis96}, \cite{LisLuc00} and this work see Sect. 2 \\
$^d$ \WCO\ from \cite{LucLis96}, \cite{LisLuc98} and this work see Sect. 2 \\
\end{table*}

\end{appendix}

\end{document}